# Optimizing the light penetration depth in APDs and SPADs for high gain-bandwidth and ultra-wide spectral response

**Ahasan Ahamed[1*], Cesar Bartolo-Perez[1], Ahmed Sulaiman Mayet[1], Soroush Ghandiparsi[1], Gerard Ariño-Estrada[2], Xiangnan Zhou[2], Julien Bec[2], Shih-Yuan Wang[3], Laura Marcu[2], M. Saif Islam[1]**

[1]Electrical and Computer Engineering Department, University of California – Davis, Davis, California, USA, 95616
[2]Department of Biomedical Engineering, University of California – Davis, Davis, California, USA, 95616
[3]W&WSens Devices, Inc., 4546 El Camino, Suite 215, Los Altos, California, USA, 94022

## ABSTRACT

Controlling light penetration depth in Avalanche Photodiodes (APDs) and Single Photon Avalanche Diodes (SPADs) play a major role in achieving high multiplication gain by delivering light near the multiplication region where the electric field is the strongest. Such control in the penetration depth for a particular wavelength of light has been previously demonstrated using integrated photon-trapping nanostructures. In this paper, we show that an optimized periodic nanostructure design can control the penetration depth for a wide range of visible and near-infrared wavelengths simultaneously. A conventional silicon APD structure suffers from high photocarrier loss due to recombination for shorter wavelengths as they are absorbed near the surface region, while silicon has low absorption efficiency for longer wavelengths. This optimized nanostructure design allows shorter wavelengths of light to penetrate deeper into the device, circumventing recombination sites while trapping the longer wavelengths in the thin silicon device by bending the vertically propagating light into horizontal modes. This manipulation of penetration depth improves the absorption in the device, increasing light sensitivity while nanostructures reduce the reflectance from the top surface. While delivery of light near the multiplication region reduces the photogenerated carrier loss and shortens transit time, leading to high multiplication gain in APDs and SPADs over a wide spectral range. These high gain APDs and SPADs will find their potential applications in Time-Of-Flight Positron Emission Tomography (TOF-PET), Fluorescence Lifetime Imaging Microscopy (FLIM), and pulse oximetry where high detection efficiency and high gain-bandwidth is required over a multitude of wavelengths.

## 1. INTRODUCTION

Avalanche Photodiodes (APDs) and Single Photon Avalanche Diodes (SPADs) are well-known in the scientific community for their capability to detect low light over a broad range of wavelengths. The high gain delivered by these detectors can enhance the signal-to-noise ratio (SNR) of faint signals, making them suitable for many biomedical applications. Biological samples limit the amount of optical input power, thus creating demands for high SNR in the detector. Fluorescence Lifetime Imaging (FLIM)[1, 2], and Time of Flight Positron Emission Tomography (TOF-PET)[3, 4] utilize the high gain-bandwidth of these devices in the visible spectrum to generate sophisticated images used for identifying cancers, malicious tumors with high precision. Pulse oximetry[5] use their high SNR (in the infrared spectrum) to detect oxygen levels in patient blood or fetus heart rate. Therefore, improving the gain-bandwidth of APDs and SPADs in the visible and near-infrared wavelengths becomes crucial for such applications.

*aahamed@ucdavis.edu; phone 530-5742762

Silicon, the most widely used semiconductor, is also the cheapest for fabricating high-performance APDs and SPADs. They also provide high sensitivity in the visible wavelengths making them useful for optical sensors and detectors used in various bio-applications. Implementation of surface nanostructures can enhance the quantum efficiency of silicon detectors to more than 60% in the near-infrared regime[6]. Recent works demonstrate significant improvement of multiplication gain by introducing micro/nanoholes into silicon APDs[7]. The speed-efficiency trade-off of the photodetector is circumvented by the engineered coupling of the vertically propagating light with photon-trapping nanostructures causing them to bend into horizontally propagating light and increasing the absorption in the material[8]. Thus, a thin detector can successfully detect a higher number of photons, improving the sensitivity of high-speed APDs and SPADs.

In our most recent findings, we observed that these nanostructures could modulate the light penetration depth in addition to improving the detection efficiency. We demonstrated that by carefully engineering the nanostructures we can deliver light to the multiplication region of the APDs or SPADs, reducing surface recombination, diffusion loss, transmission loss; thereby enhancing gain of the detectors. This control over the penetration depth enhanced the multiplication gain of APDs by more than two orders of magnitude in 450 nm wavelength[9]. Apart from single-wavelength APDs, there is a demand for high gain-bandwidth APDs over a broad range of wavelengths for reconstruction-based spectroscopy[10]. Biomedical applications such as FLIM, pulse oximetry, etc. benefit from high gain APDs throughout the visible and near-infrared spectrum.

To mitigate such demand for high gain-bandwidth, we investigated the impact of photon-trapping holes in controlling the penetration depth over a large wavelength spectrum from 300 nm to 800 nm. For a flat device, short wavelengths are absorbed near the surface, thus many of the photogenerated carriers are recombined in the surface trap sites and do not contribute to photomultiplication gain; while for longer wavelengths, most of the light transmits through the device without getting absorbed in the device. In this paper, we demonstrate that with an engineered photon-trapping nanostructure, the same device can simultaneously increase the penetration depth of short wavelengths while reducing the penetration depth of long wavelengths; thus more photogenerated carriers can participate in the impact ionization process, improving the multiplication gain of the device over the wide spectrum.

## 2. CONCEPTS

### 2.1 Photon-trapping Nanostructures
Nanostructures and nanotextures are well-known in the scientific community for improving the absorption efficiency of photodetectors and solar cells. These photon-trapping nanostructures are basically micro or nanoholes or pillars integrated on the surface of semiconductors. The incoming light interacts with these subwavelength structures and are bends horizontally into the absorbing material. This phenomenon slows down the vertical propagation of light improving the absorption in the detectors. By clever manipulation of light-matter interaction, these nanostructures can absorb more light in thinner absorbing material compared to the traditional flat devices.

### 2.2 Penetration Depth
The light intensity attenuates as it is absorbed when traveling through an absorbing material. For monochromatic light, the absorption is measured in terms of an absorption coefficient, $\alpha$. The absorption coefficient is defined as the amount of light attenuated when a unit distance is traversed in the absorbing material. In other words, the intensity of light $I(x)$ can be given by Beer-Lamberts law as $I(x) = I_0 e^{-\alpha x}$ where $I_0$ is the intensity at the surface and $x$ is the distance traversed into the absorbing material. The absorption coefficient is related to the extinction coefficient, $\kappa$ of the absorbing material. They are related by the following equation $\alpha = \frac{4\pi\kappa}{\lambda}$ where $\lambda$ is the wavelength of the monochromatic light. The absorption coefficient varies with changing wavelengths, and in practice, it reduces with increasing wavelengths.

The penetration depth, $\delta_p$ can be defined as the reciprocal of the absorption coefficient, $\alpha$, that is $\delta_p = 1/\alpha$. In other words, the penetration depth is the distance traversed by light where its intensity drops by 63.2% of its initial intensity.

Therefore, with increasing wavelength the penetration depth of light increases in silicon. However, by the addition of photon-trapping nanostructures, we can modulate the penetration depth of light over a broad spectral range to maximize APD gain (figure 1).

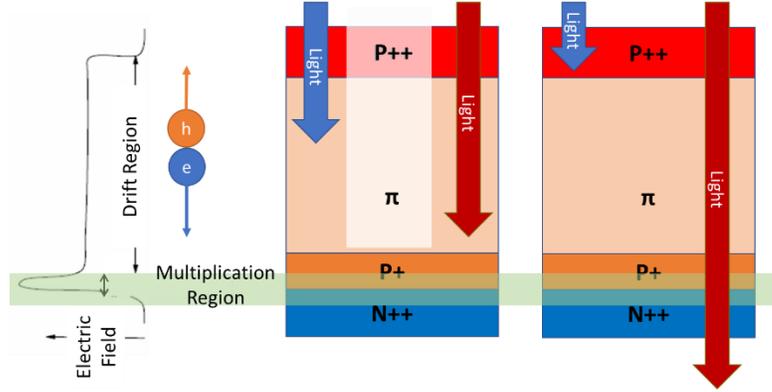

Figure 1: A schematic of a nanostructured APD compared with a conventional APD. The electric field profile based on the doping profile of the APD structure is shown on the left depicting drift and multiplication region. In nanostructured APD (center), the modulation of penetration depth allows both short (blue) and long (red) wavelengths of light to be absorbed in the intrinsic π-region consequently improving the multiplication gain. In flat APD structure (right), short (blue) wavelength is absorbed near the surface, while the long (red) wavelength transmits through the device without getting absorbed significantly.

### 2.3 Multiplication Gain

APDs and SPADs are carefully designed photodiodes to increase the multiplication gain by enhancing the impact ionization process. They have a distinct multiplication region made by very highly doped p-n junctions. The high electric field in the depletion region accelerates the free carriers increasing their momentum. Upon reaching a certain momentum they gain enough energy to generate electron-hole pairs on impact with the interstitial sites. This process creates more carriers that can again take part in the impact ionization process, creating an 'avalanche' of free carriers. The multiplication gain, $M$ is referred to the ratio of the output carriers to the input carriers. From semiconductor theory, the multiplication gain can be obtained as,

$$M = \frac{1}{1 - \int_0^L \alpha_M(x)dx} \qquad (1)$$

Here, $L$ is the depletion width and $\alpha_M$ is the multiplication coefficient for the primary carrier to be multiplied (usually electron). The multiplication coefficient is a strong function of the doping profile, applied electric field strength, and temperature. With a higher doping profile, the recombination sites also increase as well as the electric field. Therefore, to maximize the multiplication gain, the recombination rate is minimized while maintaining a high electric field to trigger avalanche.

### 3. METHODS

#### 3.1 Fabrication of Photodiode with Photon-trapping Nanoholes

In this paper, we are investigating a single p-i-n photodiode with integrated photon-trapping nanostructures. Then the results were compared with a traditional p-i-n photodiode without any nanostructures. Micro/nanoholes are etched on the surface of the detector through all the i-layer as a grating medium used for photon-trapping. A square lattice of cylindrical holes is etched into the photodiode by alternating Deep Reactive Ion Etching (DRIE) and passivation cycles. The diameter of the cylindrical hole used for this exercise is 630 nm with a periodicity of 900 nm. The detailed fabrication process can be found here[8]. Figure 2a shows the top and cross-section views of the integrated nanoholes and figure 2b shows the top view of a complete silicon photodetector with photon-trapping nanostructures and under Scanning Electron Microscope (SEM). The figures are adapted from reference[8].

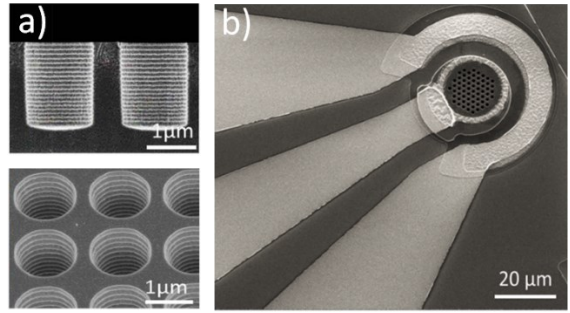

Figure 2: (a) Cross section view and top view of cylindrical nanostructures under SEM[8]. (b) Top view of a complete APD with cylindrical nanostructures under SEM[8].

### 3.2 Simulation and Analysis

For the measurement of penetration depth, we used the optical simulation tool Lumerical Finite Difference Time Domain (FDTD) simulator. To replicate the actual device conditions, we used a 2.5 μm thick silicon slab as the active absorbing layer and a low index silicon-di-oxide layer as a back-reflector to enhance the light coupling in the device. For the hole dimensions, we used 630 nm hole diameters for cylindrical-shaped holes with 900 nm periodicity in the square lattice in accordance with the fabricated device.

To calculate the absorption of light at different depths we place monitors at depths from the surface ranging from 0 μm to 2.5 μm and measure the amount of light absorbed in between two subsequent monitors. Also, monitors at the top and bottom of the device measure the reflectance and transmittance from the device, respectively. The input plane light is set from 300 nm to 800 nm wavelength range. The results are then compiled to obtain the penetration depth at different wavelengths. These results are also compared with a flat device without any photon-trapping nanostructures to show the improvements achieved.

## 4. RESULTS AND DISCUSSION

### 4.1 Reflection, Absorption, and Transmission over the visible spectrum

We simulated a flat device and a device with integrated photon-trapping structures using a Lumerical FDTD simulator for input light from 300 nm to 800 nm wavelength. The collected absorption, reflection, and transmission patterns for the flat device (figure 3a) and device with photon-trapping structures (figure 3b) are drawn against the broad range of wavelength.

We observe that for flat devices the reflection is more prominent for shorter wavelengths and there is no transmittance of light below 500 nm wavelength. However, in the case of a device with integrated nanostructures, the reflection from the top surface is significantly reduced and increased transmittance is observed for the shorter wavelength of light. This depicts that light is allowed to travel deeper in devices with nanostructures, which was not possible for a conventional device.

For longer wavelengths, we observe increased transmittance in the flat device as expected due to the low absorption coefficient of silicon at such wavelengths. But with the help of the integrated nanostructures, the incoming long wavelengths can couple into the device, allowing them to slow down and increase absorption in the active region. This hints that long wavelengths are trapped inside the thin layer of silicon.

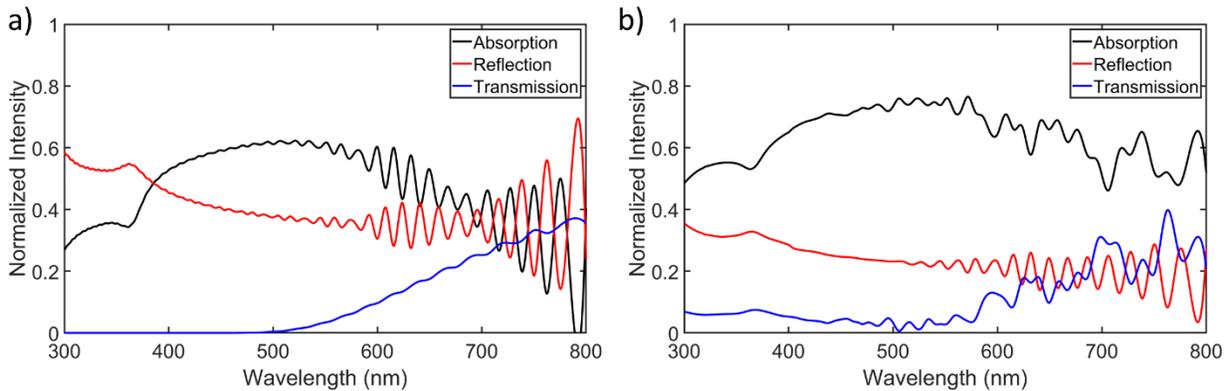

Figure 3: The absorption, reflection and transmission spectra over the wavelength range from 300 nm to 800 nm obtained from Lumerical FDTD simulation for (a) Flat device, (b) Nanostructured device with cylindrical holes with 2.5 μm thickness.

## 4.2 Absorption in layers

The light entering the semiconductor gets slowly absorbed as it propagates deeper. To understand where the light waves of different wavelengths are absorbed inside the semiconductor, we simulated the structure with monitors placed with 100 nm separations. The absorption of light in each 100 nm slab is shown in the image plot in figure 4, for flat devices (figure 4a) and devices with nanostructures (figure 4b). We can see that in the flat device, most of the short wavelengths are absorbed in the first few nanometers, while the longer wavelengths penetrate much deeper into the device, and even transmit through the 2.5 µm thick silicon structure. However, in the case of photon-trapping nanostructures, the short wavelengths can penetrate and get absorbed deeper into the device. For longer wavelengths, they interact with the nanostructures and get trapped inside the device.

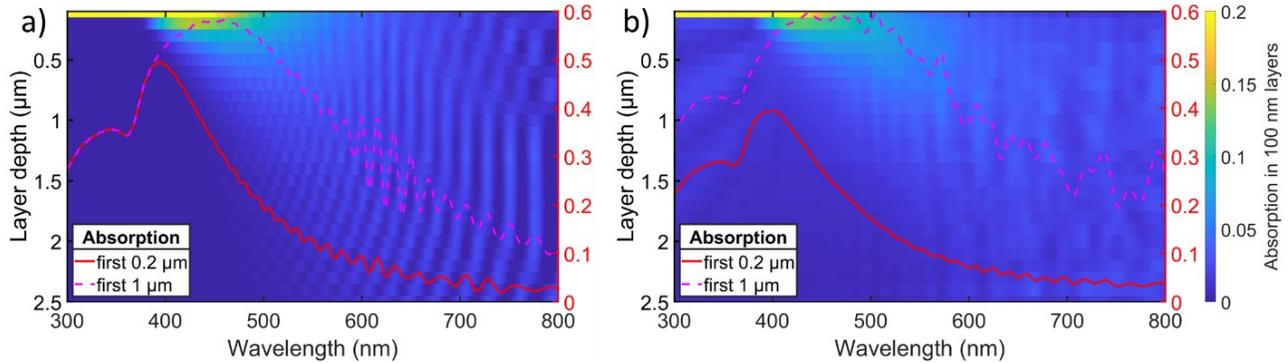

Figure 4: Lumerical FDTD simulation showing the absorption of light in each 100 nm layers over the entire depth for different wavelengths (ranging from 300 nm to 800 nm). Results are compared for (a) Flat device and (b) Nanostructured device. The amount of light absorbed in the first 0.2 µm and first 1 µm is shown in red solid line and magenta dotted line respectively. We observe that for short wavelengths, all the light is absorbed near the surface of the flat device, however, for nanostructured device they are absorbed deeper into the device. For longer wavelengths, we observe more coupling of light in the nanostructured device owing to higher absorption of 30% at 800 nm compared to 10% in flat device in the first 1 µm depth.

To get a better understanding, we also plot the absorption of light in first 0.2 µm and 1 µm in red (solid) and magenta (dotted) respectively. This shows that in a flat device all the short wavelengths are absorbed in the first 0.2 µm, but we observe lower absorption in the first 0.2 µm of the photon-trapping device and increased absorption in the first 1 µm depth. This shows that light penetrated into the deeper layers in the photon-trapping device. In the case of longer wavelengths, we observe that more light is trapped in the first 1 µm of the photon-trapping device compared to a flat device. This shows how nanostructures can modulate light penetration depth by light coupling and trapping.

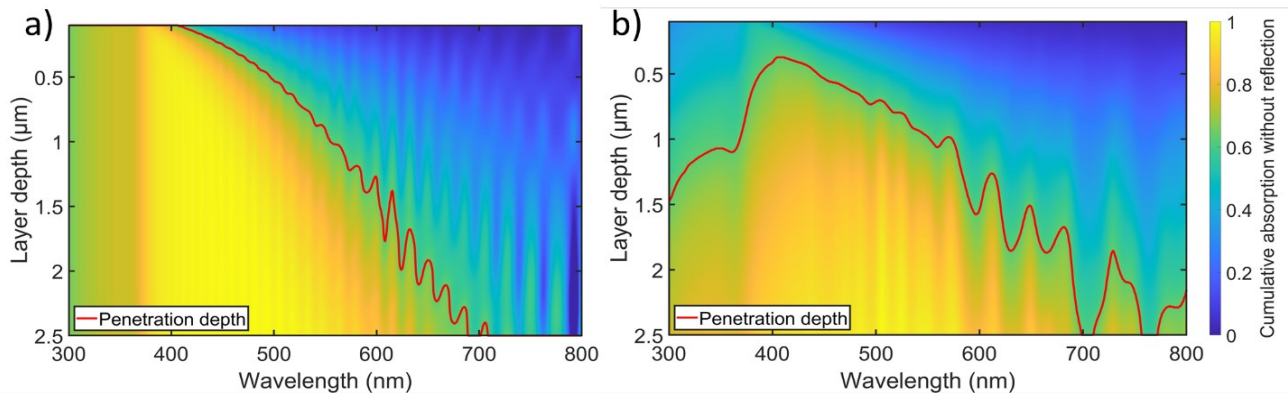

Figure 5: Lumerical FDTD simulation showing the cumulative absorption of the light of different wavelengths in (a) Flat device and (b) Nanostructured device over the depth of the absorbing material. Here the reflected light from the top surface is excluded to obtain the penetration depth (shown in red solid line) at 63.2% absorption of the light entering into the device. We observe that the penetration depth increases with increasing wavelength for a flat device (a), however in nanostructured device the penetration depth increased for shorter wavelength and decreased for longer wavelengths allowing more light to be absorbed within the device (b).

### 4.3 Penetration Depth

To obtain the penetration depth in the nanostructured device, we can not follow the conventional process using the formula $\delta_p = 1/\alpha$, where $\alpha$ is the absorption coefficient of absorbing material at a specific wavelength. This is because the absorbing material is non-homogenous, and the nanostructures can significantly perturb the propagation path of light. Therefore, we calculate the cumulative absorption of the light entering the device, excluding the reflected light from the top surface. The depth at which 63.2% of the entering light gets absorbed gives us the penetration depth for that wavelength of light.

In figure 6a-b, we show the cumulative absorption of light at different depths from the surface of the device, here the reflected light from the top surface is excluded to determine the penetration depth using the above-mentioned method. We observe that in a flat device (figure 5a), short-wavelength light is completely absorbed near the top surface of the detector, thus the penetration depth lies near the surface. With increasing wavelength, the penetration depth increases as the absorption coefficient for longer wavelength decreases. For wavelengths longer than 700 nm, the penetration depth exceeds the device depth of 2.5 μm which resonates with the poor absorption at longer wavelengths.

However, when we add photon-trapping nanostructures in the same device (figure 5b), we observe that the penetration depth increases for shorter wavelengths while reducing the penetration depth for longer wavelengths. The short-wavelength light can penetrate deeper through the hollow nanostructures (typically larger than the wavelength). But for longer wavelengths, the nanostructures are of similar order or smaller than the wavelength of light. So, they interact with the propagating light slowing them down and increasing the absorption in silicon, thus contributing to reduced penetration depth. This allows a single device to modulate the penetration depth over a broad range of wavelengths and improve the gain-bandwidth of the photodetectors.

### 4.4 Gain Measurement

This control over the penetration depth becomes extremely useful for APD structures as most of the photons will now get absorbed in the intrinsic drift region (for this device design it is situated typically from 0.5 μm to 1.8 μm). These photogenerated carriers will have a higher probability to trigger an avalanche and reduced recombination rate. Thus, the multiplication gain is enhanced by a manifold. More details on the gain improvement can be found in reference[9].

We measured the multiplication gain of the fabricated nanostructured silicon APD at 450 nm and 850 nm wavelengths and compared it with a flat silicon APD on the same wafer. Figure 6 shows the multiplication gain in these two devices where we see more than two orders of gain improvement by proper control of the penetration depth. Moreover, the same structure can perform significantly better in a broad range of wavelengths at low reverse bias voltage.

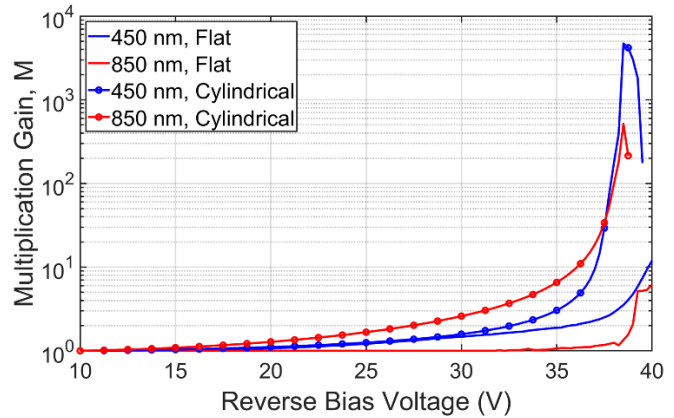

Figure 6: Comparison of multiplication gain between a nanostructured device (cylindrical holes) to a flat device at 450 nm and 850 nm wavelengths. We observe maximum multiplication gain of 4708 at 450 nm and 515 at 850 nm in the nanostructured device, where the flat device shows gain of only ~10 at both wavelengths.

### 5. CONCLUSION

We have demonstrated that with the proper implementation of photon-trapping nanostructures, we can deliver light into the drift region of the APD structures to maximize the avalanche multiplication gain of the device. Moreover, this modulation of light penetration depth can be achieved over the visible and infrared spectrum of light ensuring better gain-bandwidth crucial for applications involving cancer diagnosis, fetus pulse rate, reconstruction-based spectroscopy, etc. We have achieved more than two orders of gain improvement in the APD and SPAD at a low reverse bias voltage of only 38V. The maximum gain, M achieved in 450 nm and 850 nm are respectively 4708 and 515 in the same device, which is more than 100 times higher than that of a conventional APD structure in the same wafer.

These improved APDs and SPADs can be used in biomedical imaging modalities such as FLIM that utilizes a large portion of the visible spectra, while blood oxygen level can be determined using the near-infrared spectrum of light. Another emerging application for these devices is reconstruction-based spectroscopy which benefits from the uniqueness of the detectors. By modulating the dimensions of the nanostructures, the detectors can be made distinctly different from each other, thus could be utilized to create hyperspectral images over a broad range of wavelengths. TOF-PET imaging system can also benefit from the extremely high gain of these devices, allowing them to pinpoint the location of a malignant tumor cell with high precision.

## ACKNOWLEDGEMENTS

This work was supported in part by the Dean's Collaborative Research Award (DECOR) of UC Davis College of Engineering, the S. P. Wang and S. Y. Wang Partnership, Los Altos, CA, and by the National Institute of Biomedical Engineering and Bioengineering under the grant R21 EB028398. Cesar Bartolo-Perez acknowledges the National Council of Science and Technology (CONACYT) and UC-MEXUS for the Doctoral fellowship. Part of this study was carried out at the UC Davis Center for Nano and Micro Manufacturing (CNM2).

## REFERENCE

1. Bruschini C, Homulle H, Antolovic IM, Burri S, Charbon E. Single-photon avalanche diode imagers in biophotonics: review and outlook. *Light Sci Appl*. Published online 2019. doi:10.1038/s41377-019-0191-5
2. Unger J, Hebisch C, Phipps JE, et al. Real-time diagnosis and visualization of tumor margins in excised breast specimens using fluorescence lifetime imaging and machine learning. *Biomed Opt Express*. 2020;11(3):1216. doi:10.1364/boe.381358
3. Ariño-Estrada G, Mitchell GS, Kwon S Il, et al. Towards time-of-flight PET with a semiconductor detector. *Phys Med Biol*. 2018;63(4):4-5. doi:10.1088/1361-6560/aaaa4e
4. Gundacker S, Martinez Turtos R, Kratochwil N, et al. Experimental time resolution limits of modern SiPMs and TOF-PET detectors exploring different scintillators and Cherenkov emission. *Phys Med Biol*. 2020;65(2):025001. doi:10.1088/1361-6560/ab63b4
5. Lee H, Ko H, Lee J. Reflectance pulse oximetry: Practical issues and limitations. *ICT Express*. 2016;2(4):195-198. doi:10.1016/J.ICTE.2016.10.004
6. Cansizoglu H, Mayet AS, Ghandiparsi S, et al. Dramatically Enhanced Efficiency in Ultra-Fast Silicon MSM Photodiodes Via Light Trapping Structures. *IEEE Photonics Technol Lett*. Published online 2019. doi:10.1109/LPT.2019.2939541
7. Bartolo-Perez C, GhandiParsi S, Mayet A, et al. Controlling the photon absorption characteristics in avalanche photodetectors for high resolution biomedical imaging. In: Vol 11658. SPIE-Intl Soc Optical Eng; 2021:7. doi:10.1117/12.2577805
8. Gao Y, Cansizoglu H, Polat KG, et al. Photon-trapping microstructures enable high-speed high-efficiency silicon photodiodes. *Nat Photonics*. Published online 2017. doi:10.1038/nphoton.2017.37
9. Ahamed A, Bartolo-Perez C, Sulaiman Mayet A, et al. Controlling light penetration depth to amplify the gain in ultra-fast silicon APDs and SPADs using photon-trapping nanostructures. *https://doi.org/101117/122597835*. 2021;11800(3):118000F. doi:10.1117/12.2597835
10. Ahamad A, Ghandiparsi S, Bartolo-Perez C, et al. Smart nanophotonics silicon spectrometer array for hyperspectral imaging. In: *Conference on Lasers and Electro-Optics*. Optical Society of America; 2020:STh3M.2. http://www.osapublishing.org/abstract.cfm?URI=CLEO_SI-2020-STh3M.2